\documentclass[lineno]{JFM-FLM_Au}

\usepackage{hyperref}
\usepackage{bookmark}
\usepackage{color,bm}
\usepackage{graphicx}
\usepackage[normalem]{ulem}

\newcommand{\mean}[1]{\left\langle#1\right\rangle}

\lefttitle{V.J. Valad\~{a}o, M. Cencini, F. De Lillo, S. Musacchio and G. Boffetta}
\righttitle{Growth of small localized perturbations in SQG turbulence}

\title{Growth of small localized perturbations in Surface Quasi-Geostrophic turbulence}

\author{V.J. Valad\~{a}o\aff{1,2}, M. Cencini\aff{2,3}, F. De Lillo\aff{4}, S. Musacchio\aff{4} \and G. Boffetta\aff{4}}

\affiliation{
\aff{1}Department of Physics, University of Rome “Tor Vergata”, Via della Ricerca Scientifica 1, 00133 Rome, Italy.
\aff{2}INFN “Tor Vergata”, Via della Ricerca Scientifica 1, 00133 Rome, Italy.
\aff{3}Istituto dei Sistemi Complessi, CNR, Via dei Taurini 19, 00185 Rome, Italy.
\aff{4}Dipartimento di Fisica and INFN - Universit\`a degli  Studi di Torino, Via Pietro Giuria, 1, 10125 Torino TO, Italy.}

\corresau{V.J. Valad\~{a}o, \email{victor.de.jesus.valadao@uniroma2.it}}

\begin{document}
\maketitle
\begin{abstract}

The butterfly effect, namely the amplification of an initially localized infinitesimal perturbation, is a fundamental hallmark of chaotic systems. While this phenomenon is well understood in low-dimensional chaotic systems, its implications become more complex in spatially extended systems characterized by multiple temporal and spatial scales, such as geophysical flows. In this Letter, we investigate the butterfly effect by examining the evolution of infinitesimal localized perturbations in a minimal model of mesoscale geophysical turbulence. We find that localized perturbations exhibit highly variable dynamics, including an initial transient regime during which perturbation energy decreases while the error spectrum shifts from the injection scale toward the system’s most unstable configuration. The duration of this transient regime spans a broad range and may persist for several characteristic small-scale times, depending on the perturbation’s size, its initial location, and the Reynolds number.

\end{abstract}

\maketitle
\newpage

\section{Introduction}
\label{sec1}

Turbulence is a complex and spatio-temporally chaotic phenomenon, characterized by a huge number of interacting degrees of freedom hierarchically organized across a wide range of spatial and temporal scales \citep{frisch1995turbulence}. 
A central and longstanding question is whether turbulent flows retain any degree of predictability and, if so, how this predictability varies across scales. 
This problem traces back to the seminal works of \cite{lorenz1969predictability} and \cite{leith1972predictability}, which established that the nonlinear amplification of the initial perturbation produces  
an inverse error cascade, whereby small-scale perturbations progressively grow and propagate towards larger scales, ultimately degrading predictability across the entire flow (see \cite{boffetta1997predictability,rotunno2008generalization,boffetta2017chaos,ge2023production}).

This idea of the error cascade in turbulence has some controversial aspects nonetheless. In particular, one can expect that a small perturbation localized in the viscous scales will be dissipated before it propagates to larger scales. 
Indeed, recent works questioned the very idea of the butterfly effect in dissipative multiscale systems \citep{pielke2024butterfly}.
Surprisingly, despite the conceptual and practical relevance of the problem \citep{dubrulle2025can}, very few studies have been devoted to studying the evolution of a small and spatially localized perturbation in turbulent flows \citep{encinar2024growth,powdel2026uncertainty}.

In this Letter, we address this problem within a simple but relevant geophysical model given by the Surface Quasi-Geostrophic (SQG) equation. 
The SQG model describes the advection of the surface potential temperature in a strongly stratified and rotating environment \citep{held1995surface,lapeyre2017surface}. 
In spite of its relative simplicity, the SQG model displays a rich phenomenology with formal analogies with both 2D and 3D Navier-Stokes turbulence, as discussed in  \cite{valade2024anomalous}. 
Moreover, SQG dynamics has been used to address the predictability problem and the error cascade to large scales in several studies with the aim of generalizing the original analysis by Lorenz \citep{rotunno2008generalization,palmer2014real,durran2014atmosphere}.
By using high-resolution numerical simulations of the SQG model, we study the initial evolution of a small, localized perturbation of different sizes.
We find that during this initial phase, as the error propagates from the initial scale to the most unstable modes, the magnitude of the perturbation, measured by the energy of the difference field, can decrease in time. 
We find that the duration of this initial phase is characterized by a large variability, depending on both the initial size of the perturbation and its location in space.
In particular, we find that a small perturbation localized inside a vortical structure of the flow is strongly dissipated, leading to a decrease in the error energy for several small-scale times before experiencing the exponential chaotic growth.

\section{SQG model and numerical simulations}
\label{sec2}

The SQG model describes the large-scale dynamics of a rapidly rotating, stably stratified flow in terms of a 2D surface buoyancy field $\theta({\bf x},t)$ \citep{pierrehumbert1994spectra,lapeyre2006dynamics}
\begin{equation}
\partial_t \theta + {\bf u} \cdot {\bf \nabla} \theta = 
\nu \nabla^2 \theta + \mu \nabla^{-2} \theta + f \ ,
\label{eq1}
\end{equation}
where the incompressible velocity field ${\bf u}({\bf x},t)$ is given in terms of a stream function $\psi({\bf x},t)$ as ${\bf u}=(-\partial_y \psi,\partial_x \psi)$. 
The relation between the buoyancy field and the stream function is formally $\psi=|\nabla|^{-1} \theta$ or, in Fourier space, $\hat{\psi}=\hat{\theta}/k$. 
The dissipative terms in the SQG model represent the effects of scales
not resolved by the model, their specific form being somewhat arbitrary (see \cite{lapeyre2006dynamics}). In (\ref{eq1}) we introduce
a diffusivity $\nu$ and a large-scale damping $\mu$ chosen to be active at, respectively, small and large scales.

In the absence of forcing and dissipation, the dynamics (\ref{eq1}) conserve two invariants, the vertically integrated energy (VIE) $V=\langle \psi \theta \rangle_x/2$, which is the SQG analogue of the total energy of the associated three-dimensional quasi-geostrophic flow, and the surface potential energy (SPE) $E=\langle \theta^2 \rangle_x/2$, which measures the variance of the surface buoyancy field.
Brackets $\langle ...\rangle_{x}$ indicate the average over the physical domain. A peculiarity of the SQG model is that, because of the relation between surface buoyancy and velocity, the SPE is also equal to the surface kinetic energy, i.e. $\langle \theta^2 \rangle_x = \langle |\mathbf{u}|^2 \rangle_x$.

In forced-dissipative SQG turbulence, these two inviscid invariants are transferred in opposite directions.
The SPE undergoes a direct cascade towards small scales, where it is removed by the diffusive term $\nu\nabla^2\theta$.
The VIE undergoes an inverse cascade towards large scales, where it is removed by the large-scale friction term $\mu\nabla^{-2}\theta$.
The viscous and friction terms therefore provide, respectively, the small-scale and large-scale cutoffs required to obtain a statistically stationary state.

In the turbulent regime, the SQG flow produces a direct cascade of SPE {\it \`a la} Kolmogorov with an SPE spectrum, or energy spectrum close to the dimensional prediction \citep{pierrehumbert1994spectra}
\begin{equation}
E(k) \simeq \varepsilon^{2/3} k^{-5/3}
\label{eq2}
\end{equation}
in the range of wavenumbers $1/\ell_f \ll k \ll 1/\ell_{\nu}$ where $k=|{\bf k}|$, $\ell_{\nu}=(\nu^3/\varepsilon)^{1/4}$ is the diffusive (Kolmogorov) small scale and $\varepsilon$ is the average SPE viscous dissipation rate, $\varepsilon=\nu\mean{|\nabla\theta|^2}_{x,t}$.

We remark that the scaling (\ref{eq2}) is not exact, as the direct cascade in SQG 
displays intermittency corrections, similarly to 3D Navier-Stokes turbulence \citep{valade2025surface}. 
Since the effects of these corrections are relatively small for low-order observables, such as the energy spectrum, they will not be considered in this study.

We perform numerical simulations of (\ref{eq1}) in a square box of size $L=2\pi$ with periodic boundary conditions at resolution $8192^2$ and three different Reynolds numbers $Re=\varepsilon^{1/3}\ell_f^{4/3}/\nu$ (see \cite{valadao2025spectrum} for details on the implementation of the code). 
In all the simulations, the dissipative scale is well resolved as the maximum wavenumber $k_{\max}$ satisfies $k_{\max} \ell_{\nu} \ge 3.0$.
After a transient, the system reaches a statistically  stationary turbulent state characterized by a well-developed Kolmogorov spectrum (\ref{eq2}). 
In these conditions, we computed the leading Lyapunov exponent $\lambda$ by measuring the rate of logarithmic divergence of two close realizations averaged over a very long trajectory, a standard method in the study of dynamical systems \citep{cencini2009chaos}. 
For the simulations at different Reynolds numbers, see Table \ref{table1}, which reports the main properties of the turbulent flow (see \cite{valadao2025scaling} for more details). 

\begin{table}
\begin{center}
\begin{tabular}{c|cccccc}
Run & $Re$ & $E$ & $\ell_{\nu}$ & $\tau_{\nu}$ & $\lambda$ &  $\varepsilon$ \\
\hline 
$1$ & $15900$ & $ 49.5 $ & $0.0016$ & $0.0062$ & $7.32$ & $10.41$\\
$2$ & $21200$ & $ 49.9 $ & $0.0013$ & $0.0054$ & $8.97$ & $10.20$\\
$3$ & $25400$ & $ 50.3 $ & $0.0011$ & $0.0049$ & $10.2$ & $10.17$
\end{tabular}
\caption{
Parameters of the simulations. The forcing scale is kept constant throughout the simulations:  $\ell_f\approx 1.8$. The typical timescale at the forcing scale is given by $\tau_f=\varepsilon^{-1/3}\ell_f^{2/3}\approx0.68$. The amplitude $\delta_0$ is fixed by the requirement on the initial error energy $E_\Delta(0)=5\times10^{-9}$.}
\label{table1}
\end{center}
\end{table}

In the stationary turbulent conditions, at time $t=t_0$ we locally modify the reference buoyancy field $\theta({\bf x},t_0)$ around the point ${\bf x}_0$ to generate a perturbed field $\theta'({\bf x},t_0)=\theta({\bf x},t_0)+\sqrt{2}\ \delta \theta_0({\bf x}-{\bf x}_0)$.
We remark that we consider a zero-average buoyancy field and therefore also the perturbation has a vanishing average over the domain $\langle \delta \theta_0 \rangle_x=0$. 
Because of the non-local nature of the model (\ref{eq1}) (due to the incompressibility of the velocity field), a generic perturbation localized in $\theta$ does not correspond to a localized perturbation in the velocity field, producing an almost instantaneous spread of the perturbed field. 
Nonetheless, it is possible to produce a perturbation localized both in the scalar and velocity fields by the combination of a Gaussian with an algebraic decay in Fourier space.
Written for the stream function, the perturbation has the form  
$\delta \hat{\psi}_0({\bf k};{\bf x}_0)=\delta_0 e^{-\sigma^2 k^2 / 2} k^{-\beta} e^{-i{\bf k}\cdot{\bf x}_0}$, for $\bf{k}\neq0$, and zero otherwise. We fix the exponent $\beta=0.4$ from the request that the perturbations in the velocity and in the scalar fields have the same large-distance asymptotic behavior (see Appendix~\ref{appendix1}).
The amplitude $\delta_0$ is sufficiently small to guarantee the evolution of the perturbation in the linear regime. 
The width $\sigma$ of the perturbation is the control parameter that we change, together with the Reynolds number, to investigate the effect of spatial confinement on the evolution of the error. 

The two simulations of the reference and perturbed fields are then evolved (with the same realization of the random forcing) and we compute the time evolution of the difference field $\delta \theta({\bf x},t)=(\theta'({\bf x},t)-\theta({\bf x},t))/\sqrt{2}$ whose dynamics, in the linear regime, is ruled by 
\begin{equation}
\partial_t \delta \theta + {\bf u} \cdot {\bf \nabla} \delta \theta =
\nu \nabla^2 \delta \theta + \mu \nabla^{-2} \delta \theta 
- \delta {\bf u} \cdot {\bf \nabla} \theta \, .
\label{eq3}
\end{equation}

The distance between the two realizations of the turbulent flow is quantified by the error energy $E_{\Delta}$ and the error energy spectrum $E_{\Delta}(k,t)$  
\begin{equation}
E_{\Delta}(t) = \int E_{\Delta}(k,t) dk = {1 \over 2} \langle \delta \theta^2({\bf x},t) \rangle_x \ ,\
\label{eq4}
\end{equation}
which, according to (\ref{eq3}), evolves as 
\begin{equation}
{d E_{\Delta} \over dt} = -\nu \langle (\nabla \delta \theta)^2 \rangle_x
- \mu \langle (\nabla^{-1} \delta \theta)^2 \rangle_x 
- \langle \delta \theta \delta {\bf u} \cdot {\bf \nabla} \theta \rangle_x \ .\
\label{eq5}
\end{equation}
We note that with the above definition for the difference field, one has $E_{\Delta}=E$ for completely uncorrelated fields $\theta$ and $\theta'$. 
The balance equation (\ref{eq5}) shows that the last term in (\ref{eq3}) is the one responsible for the error growth, while the dissipative terms, at large and small scales, simply reduce the error.

\begin{figure}[h!]
\centering
\includegraphics[width=0.49\textwidth]{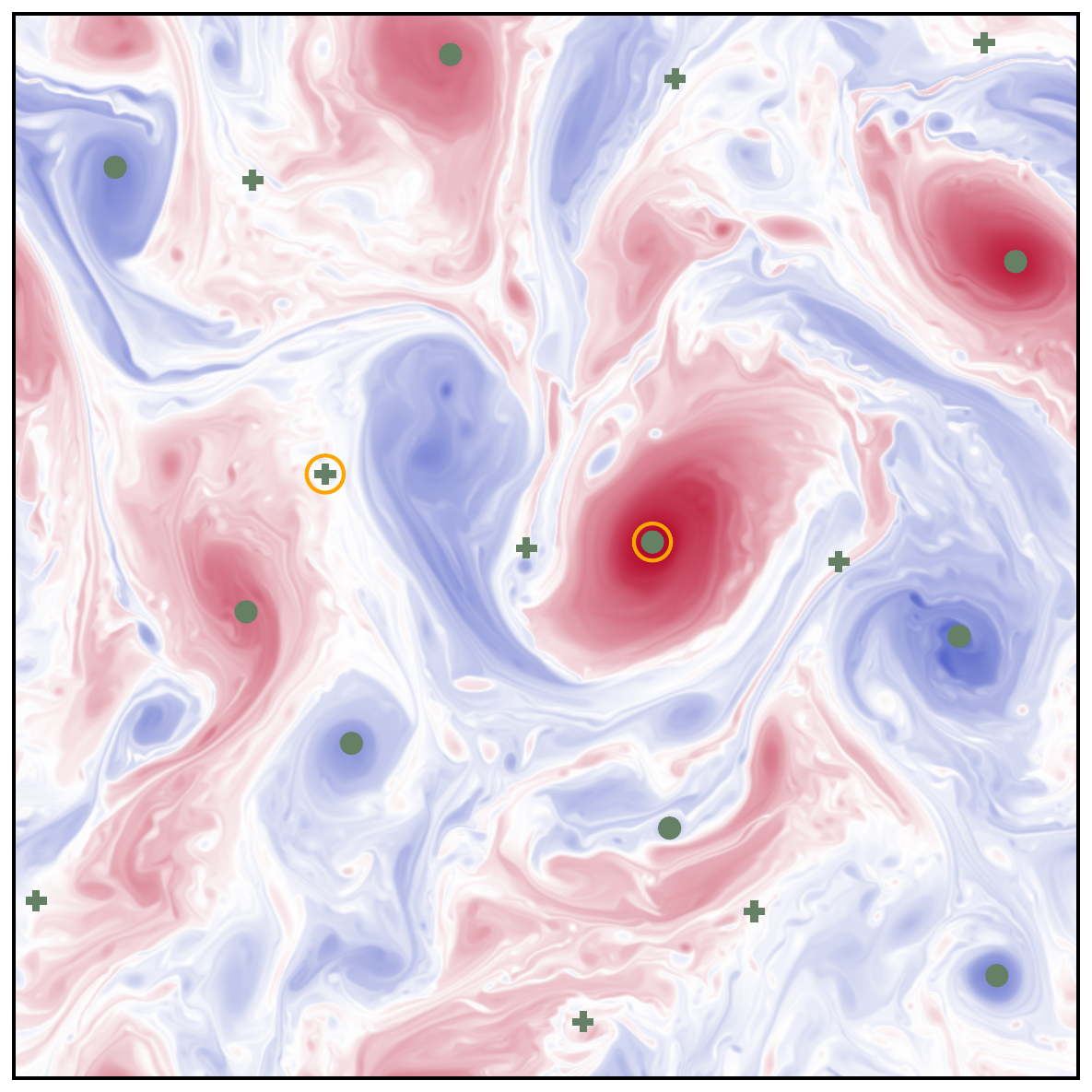}
\includegraphics[width=0.49\textwidth]{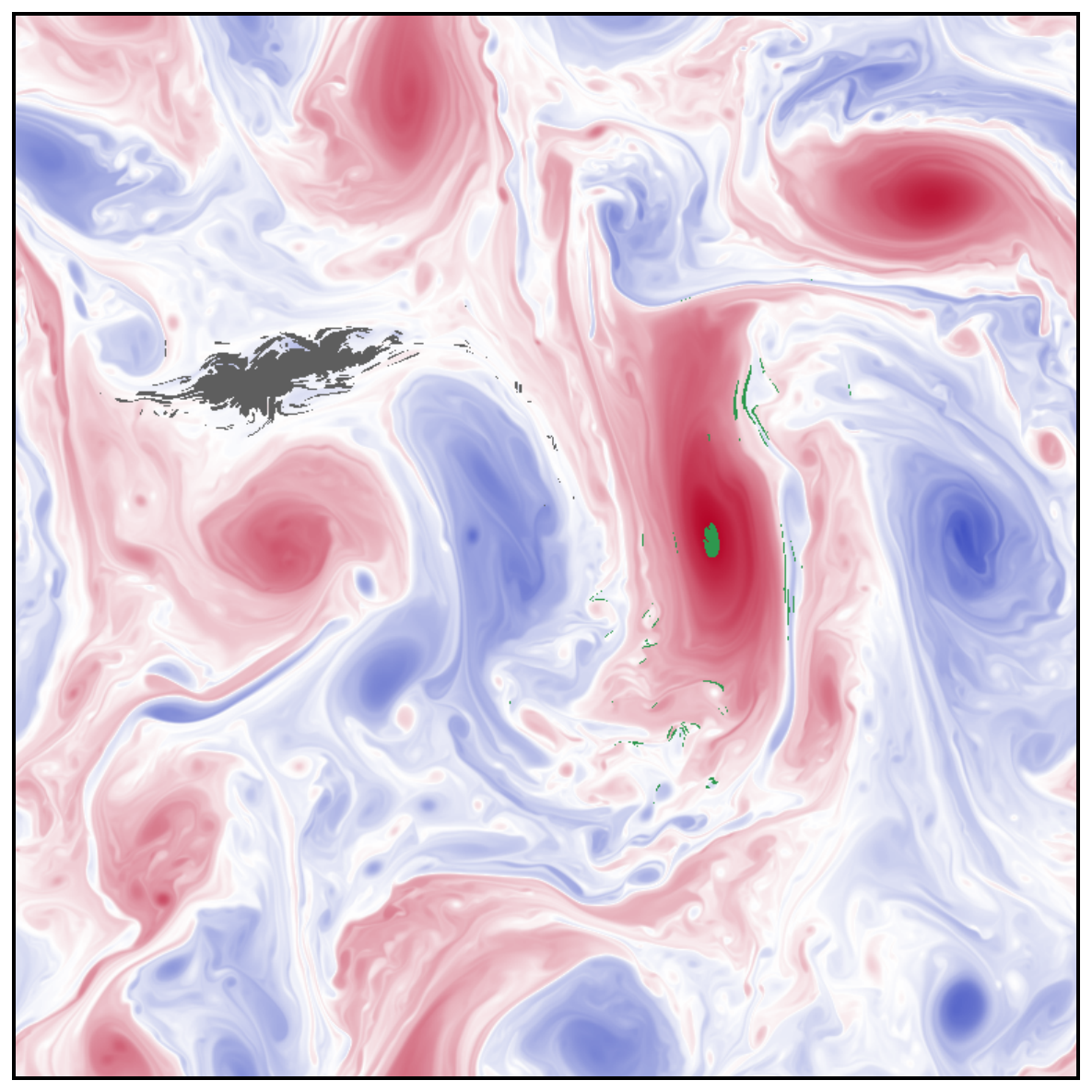}
\caption{
Left panel: Initial turbulent scalar field at $Re=25400$ with the positions of the $18$ initial perturbations located by the maximum (circles) and minimum (crosses) of $|\theta({\bf x})|$ in a $3 \times 3$ division of the domain. 
Right panel: spreading of the perturbation at time $t \lambda=2$ in the two cases indicated by an orange circle in the left panel. 
The regions where the perturbation exceeds the threshold $\delta_0=\sqrt{2E_\Delta(0)}=10^{-4}$ are shown in black/green.
}
\label{fig1}
\end{figure}
Figure~\ref{fig1} (left panel) shows a snapshot of a turbulent field $\theta({\bf x},t)$ together with the location ${\bf x}_0$
of $18$ different initial perturbations. To test the variability of local perturbations, we divide the computational box into a $3 \times 3$ grid. In each subdomain of the grid, we locate the maximum and minimum of $|\theta({\bf x})|$ and we initialize the perturbation on these points. As shown in Fig.~\ref{fig1}, maxima (represented by filled circles) are at the center of vortices, while minima (plus symbols) are in the background field between vortices. For each perturbation, we compute the evolution of the difference $\delta \theta({\bf x},t)$. The right panel of Fig.~\ref{fig1} shows the dispersion of the difference field (plotted above a threshold) after a dimensionless time $t \lambda=2$ for the two cases labeled with an orange circle in the 
left panel. It is evident that, when the perturbation is injected inside a vortex, it remains confined and small, while the spreading is much more effective in the case outside the vortex.

\begin{figure}[h!]
\centering
\includegraphics[width=0.48\textwidth]{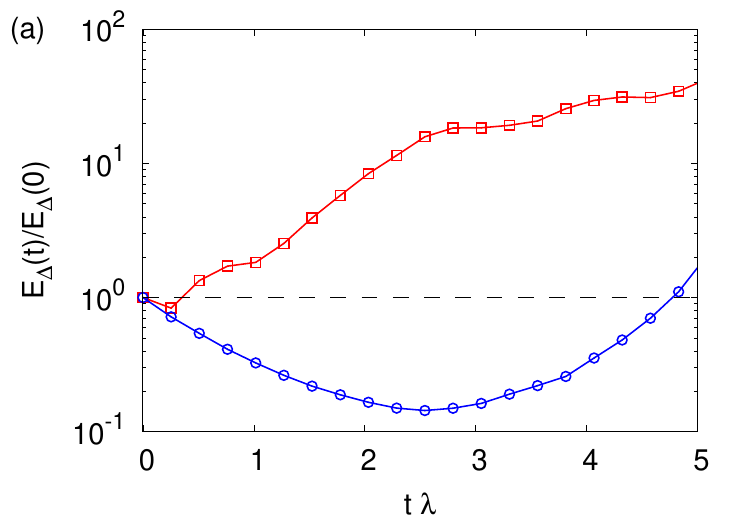}
\includegraphics[width=0.48\textwidth]{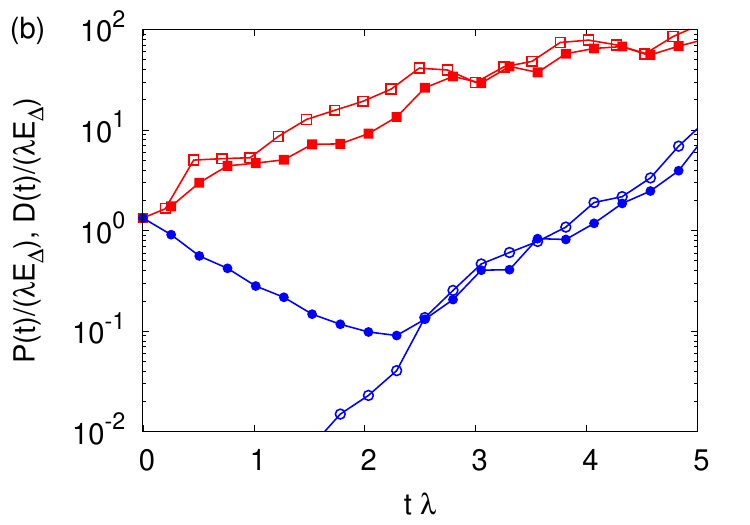}
\caption{(a) Time evolution of the error energy $E_{\Delta}$ normalized with the initial error for the two cases highlighted in orange in Fig.~\ref{fig1}. Blue circles correspond to the case with the initial perturbation inside a vortex (highlighted circle in Fig.~\ref{fig1}), red squares to the initial perturbation with minimum $|\theta({\bf x}_0)|$ (highlighted cross in Fig.~\ref{fig1}).
(b) Error energy production term $P(t)=-\langle \delta \theta \delta {\bf u} \cdot {\bf \nabla} \theta \rangle_x$ (open symbols) and viscous dissipative term $\nu \langle ({\bf \nabla} \delta \theta)^2 \rangle_x$ (filled symbols) normalized with $\lambda E_{\Delta}(0)$. Blue circles and red squares correspond to the two cases in panel (a).}
\label{fig2}
\end{figure}
The time evolution of the error energy for these two cases is shown in Fig.~\ref{fig2}a.
In both cases, we observe an initial decrease of $E_{\Delta}$ due to the dissipation of the perturbation by the diffusive terms (we remark that, by construction, the production term at $t=0$ vanishes as shown in Appendix~\ref{appendix1}). 
After a transient phase, the error energy starts to increase exponentially as expected in a chaotic system. Remarkably, we find that the initial decaying phase displays a large variability, which can last several small-scale times, as shown by these two examples (see Fig.~\ref{fig:energytrajectories} in Appendix~\ref{appendix2} for the evolution of $E_{\Delta}$ on the whole set of 18 simulations).

To better understand the source of this variability, in Fig.~\ref{fig2}b we plot the time evolution of the two contributions to the error energy balance, i.e., the dissipation and the production terms in the RHS of (\ref{eq5}). 
We see that the production term remains very small for a long time when the perturbation is inside a vortex. 
This can be understood geometrically from the fact that in an almost circular vortex, $\nabla \theta$ is predominantly radial, while both the perturbation velocity $\delta{\bf u}$ and the velocity field ${\bf u}$ are initially oriented along the azimuthal direction. This local misalignment between $\nabla \theta$ and $\delta {\bf u}$ suppresses the initial evolution of the production term in (\ref{eq5}), while the advection term ${\bf u}\cdot\nabla\delta\theta$ is also reduced. As a result, dissipation dominates the dynamics over a large time interval, delaying the onset of exponential growth.

\begin{figure}[h!]
\centering
\includegraphics[width=0.48\textwidth]{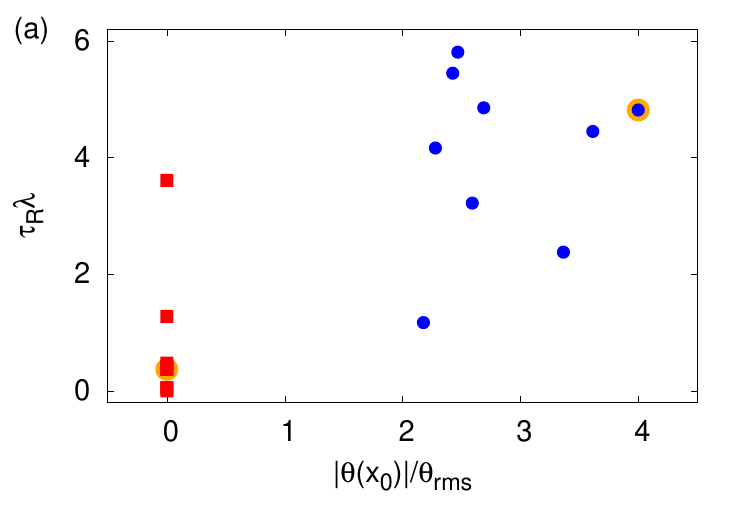}
\includegraphics[width=0.48\textwidth]{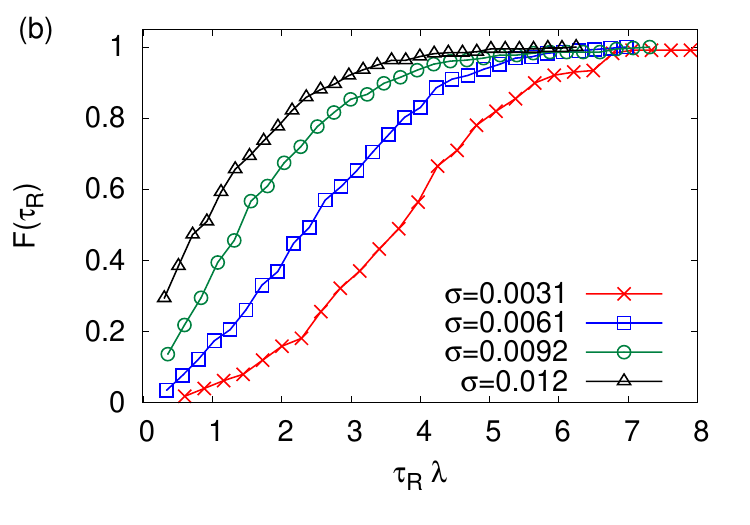}
\caption{(a) Return time as a function of the initial magnitude of the buoyancy field at the perturbation point $|\theta({\bf x}_0)|$ for the $18$ simulation of Fig.~\ref{fig1}.
(b) Cumulative distribution function $F(\tau_R)$ of the return time $\tau_R$ for the simulation at $Re=15900$ and a different initial size of the error. Distributions are computed on many realizations ranging from $190$ to $375$ from smaller to larger $\sigma$, respectively.}
\label{fig3}
\end{figure}

The observed variability calls for a statistical analysis of the initial decaying phase.
To this aim, we define the return time $\tau_R$ as the time required for the error energy to recover its initial value, $E_{\Delta}(\tau_R)=E_{\Delta}(0)$, and we also compute the minimum value $E_{\min}$ reached by $E_{\Delta}$ during this time interval.
Figure~\ref{fig3}a shows the return time for the $18$ simulations of Fig.~\ref{fig1}a as a function of the magnitude of the initial scalar field.
A clear tendency toward two groups is observed, associated with perturbations initialized inside (large $|\theta(\mathbf{x}_0)|$, circles) or outside (small $|\theta(\mathbf{x}_0)|$, squares) coherent vortical structures.
In most cases, a small (large) initial value of $|\theta(\mathbf{x}_0)|/\theta_{\rm rms}$ is associated with a short (long) return time, as in the examples of Fig.~\ref{fig2}.
Nonetheless, there are a few cases in which this simple correlation fails.
This is because $\tau_R$ depends not only on the initial local structure of the flow, but also on its subsequent evolution.
For instance, in two of the $18$ realizations, the vortex initially surrounding the perturbation is rapidly dissipated by the dynamics.
Therefore, although $|\theta(\mathbf{x}_0)|$ provides a useful indicator of the return time, the fate of a localized perturbation cannot in general be inferred from its initial environment alone.

In order to characterize the transient behavior of the perturbation, we performed a statistical 
analysis with many (between $200$ and $300$) realizations of the initial perturbation placed at random initial positions in the domain. The numerical experiment is then repeated for different sizes $\sigma$ of the initial perturbation and for $3$ different Reynolds numbers. 

An example of the statistics obtained from these numerical experiments for the case at lower Re is shown in Fig.~\ref{fig3}, where we plot the cumulative distribution function of the return times $F(\tau_R)$ for four different values of the perturbation size $\sigma$. 
It is evident that the return time ranges from a small fraction of $\lambda$ to several Lyapunov times and that this variability is more extreme for small perturbations. 
Moreover, the distribution is shifted to larger values of $\tau_R$ for a smaller size of the initial error. 

\begin{figure}[h!]
\centering
\includegraphics[width=0.48\textwidth]{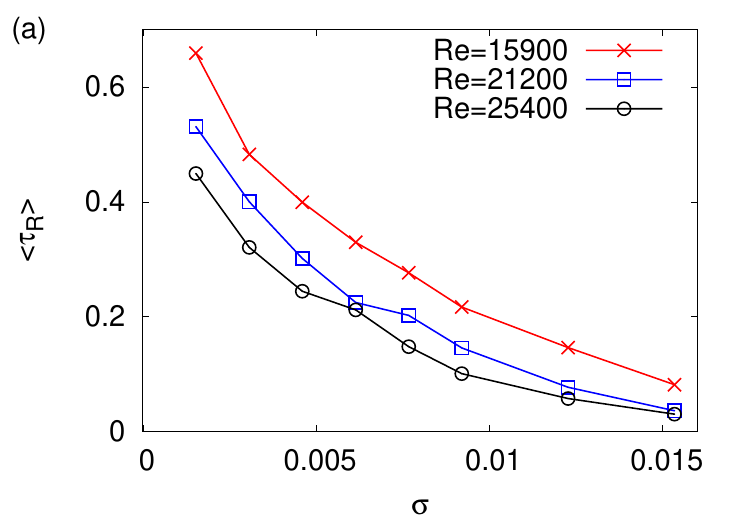}
\includegraphics[width=0.48\textwidth]{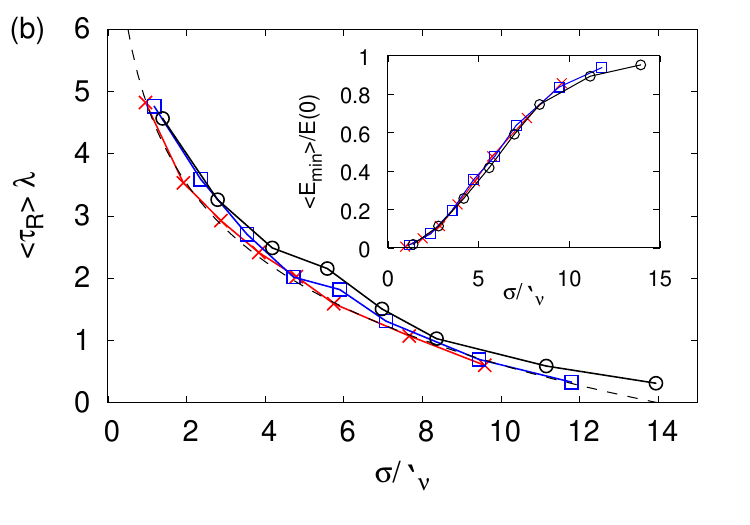}
\caption{a) Mean return time $\langle \tau_R \rangle$ as a function of the perturbation width $\sigma$ for simulations at different Reynolds numbers. b) Mean return time rescaled with the Lyapunov exponent $\lambda$ as a function of the perturbation width $\sigma$ made dimensionless with $\ell_{\nu}$. Colors and symbols as in a). The dashed line represents $-1.81 \log(\sigma/\ell_{\nu})+4.78$. Inset: the minimum error energy normalized with the initial one as a function of perturbation width.  }
\label{fig4}
\end{figure}

From the distribution shown in Fig.~\ref{fig3}, we compute the mean value of the return times $\langle \tau_R \rangle$, plotted in Fig.~\ref{fig4}a as a function of the perturbation width $\sigma$ and for the different Reynolds numbers. 
Figure~\ref{fig4}a shows that the return time increases for smaller $\sigma$, as expected from Fig.~\ref{fig3}, and decreases by increasing the Reynolds number, i.e., the Lyapunov exponent, of the flow.

\begin{figure}[h!]
\centering
\includegraphics[width=0.52\textwidth]{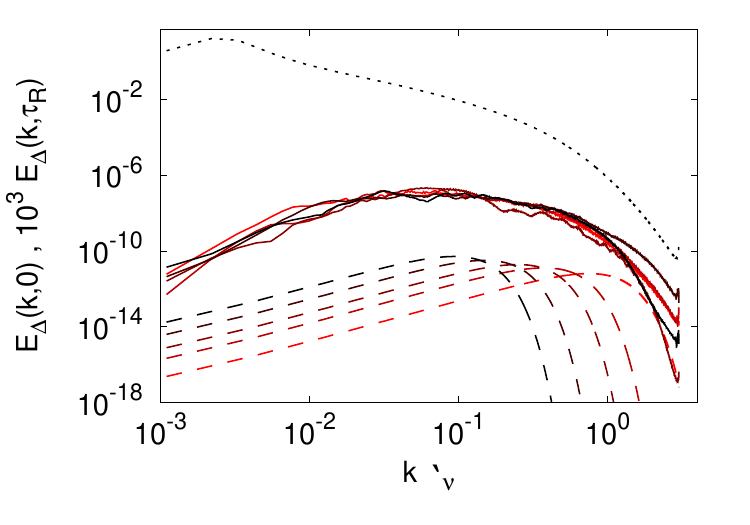}
\caption{Initial error energy spectra $E_{\Delta}(k,0)$ (dashed lines) and error energy spectra at the final time $E_{\Delta}(k,T_f)$ (continuous line) for initial error sizes $\sigma=1.4 \ell_\nu$, $\sigma=2.8\ell_\nu$, $\sigma=4.2\ell_\nu$, $\sigma=7.0\ell_\nu$, $\sigma=10.9\ell_\nu$ (from red to black) for Run 3. For clarity, the energy spectra at the final time are multiplied by a factor $10^3$. The black dotted line represents the energy spectrum $E(k)$ of the field. 
}
\label{fig5}
\end{figure}

In order to understand the dependence of $\langle \tau_R \rangle$ on $\sigma$, we consider the error spectra in Fourier space. 
Figure~\ref{fig5} shows the evolution of the error spectra $E_{\Delta}(k,t)$ for the simulation at $Re=25400$. Initial perturbations at different values of $\sigma$ correspond to different initial spectra peaked at wavenumbers $k_{\sigma} \propto 1/\sigma$, shown as dashed lines in Fig.~\ref{fig5}.
At the time $T_f$, when the error energy reaches the value $E_{\Delta}(T_f)=4 E_{\Delta}(0)$, the error growth is in the exponential regime and the error energy spectrum assumes a universal shape peaked at the most unstable wavenumber $k_{u}$, independent of the initial value of $\sigma$. 
We remark that, as clear from Fig.~\ref{fig5}, we are still in the linear regime in which the error field grows according to (\ref{eq3}).
At longer times, not investigated here, we enter the non-linear regime in which the error is transferred to larger scales and eventually saturates. 

The evolution of the error in Fourier space shown in Fig.~\ref{fig5} allows us to understand the dependence of $\langle \tau_R \rangle$ on $\sigma$ by a simple argument based on two modes involved in the process. 
The first is $k_{\sigma}$, which is the most energetic mode determined by the initial perturbation. 
The second is the most unstable mode $k_u$, which is known to be around
$k_u \ell_{\nu} \simeq 0.1$ \citep{valadao2025scaling}.
This is the wavenumber at which the error energy spectrum $E_{\Delta}(k)$ grows exponentially with the Lyapunov exponent of the flow, i.e. $E_{\Delta}(k_u,t) \simeq E_{\Delta}(k_u,0) e^{2 \lambda t}$ (for simplicity we neglect the fluctuations of the Lyapunov exponent, see e.g. \cite{crisanti1988generalized}).

For the perturbation with spatial decaying exponent $\beta$ concentrated at wavenumber $k_{\sigma} \sim 1/\sigma \gg k_u$, the initial ratio of the energies at the two modes is $E_{\Delta}(k_u,0)/E_{\Delta}(k_{\sigma},0) \simeq (k_u/k_{\sigma})^{3-2 \beta} \ll 1$.
Since $k_{\sigma}$ is in the dissipative range, the corresponding energy is strongly dissipated while the energy at $k_u$ grows exponentially.
Therefore, the time it takes for the error energy at wavenumber $k_u$ to reach the initial value can be estimated from the condition $(k_u/k_{\sigma})^{3-2 \beta} e^{2 \lambda \tau_R} = 1$ which means $\tau_R \simeq -(C/\lambda)\log(\sigma/\ell_\nu)$ where $C$ is a constant which depends on $\beta$. Fig.~\ref{fig4}b shows that this prediction works very well when we rescale the mean return time with the Lyapunov exponent of the flow and $\sigma$ with the dissipative scale $\ell_\nu$. The inset of Fig.~\ref{fig4}b shows the minimum value $E_{min}$ reached by the error energy normalized with the initial error. When $\sigma/\ell_\nu \ge 10$ the minimum disappears since $\tau_R \to 0$. 

\section*{Discussion and Conclusions}

We studied the butterfly effect in the forced-dissipative SQG turbulence, a classical two-dimensional model for geophysical flows which has many analogies with the three-dimensional Navier--Stokes equations. 
By perturbing the turbulent flow with a small perturbation localized in space, we find that exponential amplification of the perturbation is preceded by a pre-asymptotic stage, during which the spectrum of the perturbation aligns with the most unstable wavenumber and the energy of the error field decreases in time. 

We study the statistics of this initial phase, and we find that its duration, the return time, is characterized by a large variability depending on the local properties of the turbulent flow at the position of the initial perturbation.
The average return time depends logarithmically on the scale of the perturbation, and this dependence can be heuristically understood as the result of a competition between two scales: the injection scale of the perturbation, which decays, and the most unstable scale of the flow, which grows exponentially. 

Space-time intermittency is associated with the long-lived large-scale coherent structures observed in SQG turbulence \citep{valadao2025scaling,valadao2024nonequilibrium}. 
We have seen that when the initial perturbation is localized inside a vortex, the return time is typically large, but there are exceptions, and the fate of the perturbation does not depend only on the local magnitude of the scalar field. A more complete characterization of the conditions necessary to achieve long predictability with localized perturbations is left for future investigations.
We also remark that the observed transient behavior depends on the norm used to measure the trajectory separation. While the energy norm (\ref{eq4}) is a natural one, other choices are possible, and we expect that the quantitative behavior observed during the initial transient depends on the norm chosen. 

The central question behind the real butterfly effect, i.e., whether a small localized perturbation is ultimately absorbed by the dissipative nature of real turbulent flows, cannot ignore the role of small-scale noise \citep{thalabard2020butterfly,bandak2024spontaneous}.
From a conceptual point of view, when a small perturbation in the dissipative range initially decays, it can reach very small scales and amplitudes where the role of thermal noise cannot be neglected. 
While this effect is irrelevant in the present SQG model—where viscous dissipation represents a parametrization of unresolved smaller scales—it is expected to play a role in 3D Navier–Stokes turbulence, where molecular viscosity is explicitly resolved.
It would therefore be very interesting to study the effects of thermal noise on an initially small perturbation in conditions accessible to laboratory experiments.

\section*{Acknowledgement}

We acknowledge HPC CINECA for computing resources within the grants INFN25-FieldTurb and INFN26-FieldTurb. We are indebted to Angelo Vulpiani for the inspiration behind this work. V. J. V. acknowledges the support by the Italian Ministry of University and Research (MUR) - Fondo Italiano per la Scienza (FIS2) - 2023 Call, project DeepFL, CUP: E53C24003760001.

\bibliographystyle{jfm}
\bibliography{biblio}

\clearpage
\appendix
\makeatletter
\renewcommand{\theHsection}{A.\arabic{section}}
\makeatother

\section{The localized perturbation}
\label{appendix1}

\ \ \ In this appendix, we discuss the shape of the localized perturbation used in the simulations. 
For simplicity, we set ${\bf x}_0=0$; the dependence on ${\bf x}_0$ is recovered by the phase factor $e^{-i{\bf k}\cdot{\bf x}_0}$ in Fourier space. 
The perturbation is prescribed for the stream function as
\begin{equation}
\delta\hat\psi_0(k)
=
A e^{-\sigma^2 k^2/2} k^{-\beta}.
\end{equation}

The perturbation is radially symmetric in physical space. Therefore, its inverse Fourier transforms can be written as Hankel transforms. 
Thus, we can write the stream function as
\begin{equation}
\delta\psi_0(r)
=
\frac{A}{2\pi}
\int_0^\infty
k^{1-\beta}
e^{-\sigma^2 k^2/2}
J_0(kr)\,dk ,
\label{app:eq:psi_hankel}
\end{equation}
where $r=|{\bf x}|$, and the buoyancy field is given by
\begin{equation}
\delta\theta_0(r)
=
\frac{A}{2\pi}
\int_0^\infty
k^{2-\beta}
e^{-\sigma^2 k^2/2}
J_0(kr)\,dk .
\label{app:eq:theta_hankel}
\end{equation}
The induced velocity perturbation is the divergenceless field given by
\begin{equation}
\delta{\bf v}_0=(-\partial_y ,\partial_x )\delta\psi_0 .
\end{equation}
Since $\delta\psi_0$ is radial, $\delta{\bf v}_0$ is azimuthal,
\begin{equation}
\delta{\bf u}_0(r)
=
\frac{d\delta\psi_0}{dr}\hat{\bf e}_\varphi .
\end{equation}
At this point, it is then clear that the production term at the initial time $P(0)=-\mean{\delta\theta_0 \delta{\bf u}_0\cdot\nabla\theta}$ vanishes independently of the underlying $\theta$ due to the orthogonality of $\delta{\bf u}_0$ and $\nabla\delta \theta_0$. Using $J_0'(z)=-J_1(z)$, one obtains
\begin{equation}
\delta{\bf u}_0(r)
=
-\frac{A}{2\pi}
\left[
\int_0^\infty
k^{2-\beta}
e^{-\sigma^2 k^2/2}
J_1(kr)\,dk
\right]\hat{\bf e}_\varphi .
\label{app:eq:v_hankel}
\end{equation}
All the integrals in the definitions of the perturbation buoyancy, velocity, and stream functions are particular cases of
\begin{equation}
I_{\mu,\nu}(r,\sigma)
\equiv
\int_0^\infty
k^\mu e^{-\sigma^2 k^2/2}J_\nu(kr)\,dk ,
\end{equation}
which can be written as
\begin{equation}
I_{\mu,\nu}(r,\sigma)
=
\frac{r^\nu}{2^{\nu+1}}
\left(\frac{2}{\sigma^2}\right)^{(\mu+\nu+1)/2}
\frac{\Gamma\left(\frac{\mu+\nu+1}{2}\right)}
{\Gamma(\nu+1)}
\,{}_1F_1\left(
\frac{\mu+\nu+1}{2};
\nu+1;
-\frac{r^2}{2\sigma^2}
\right).
\label{app:eq:general_integral}
\end{equation}
Applying (\ref{app:eq:general_integral}) to (\ref{app:eq:psi_hankel}), one has
\begin{equation}
\delta\psi_0(r)
=
\frac{A}{2\pi}
2^{-\beta/2}
\sigma^{\beta-2}
\Gamma\left(1-\frac{\beta}{2}\right)
\,{}_1F_1\left(
1-\frac{\beta}{2};
1;
-\frac{r^2}{2\sigma^2}
\right).
\label{app:eq:psi_hyper}
\end{equation}
Similarly,
\begin{equation}
\delta\theta_0(r)
=
\frac{A}{2\pi}
2^{(1-\beta)/2}
\sigma^{\beta-3}
\Gamma\left(\frac{3-\beta}{2}\right)
\,{}_1F_1\left(
\frac{3-\beta}{2};
1;
-\frac{r^2}{2\sigma^2}
\right),
\label{app:eq:theta_hyper}
\end{equation}
and
\begin{equation}
\delta{\bf u}_0(r)
=
-\frac{A}{2\pi}
2^{-\beta/2}
r\sigma^{\beta-4}
\Gamma\left(2-\frac{\beta}{2}\right)
\,{}_1F_1\left(
2-\frac{\beta}{2};
2;
-\frac{r^2}{2\sigma^2}
\right)
\hat{\bf e}_\varphi .
\label{app:eq:v_hyper}
\end{equation}

The far-field behavior follows from the asymptotic expansion
\begin{equation}
{}_1F_1(a;c;-z)
\simeq
\frac{\Gamma(c)}{\Gamma(c-a)}z^{-a},
\qquad
z\to+\infty .
\label{app:eq:hyper_asymptotic}
\end{equation}
Using $z=r^2/(2\sigma^2)$, one obtains, for $r\gg\sigma$,
\begin{equation}
\delta\theta_0(r)
\simeq
C_\theta(\beta)A r^{\beta-3},
\qquad
\delta{\bf u}_0(r)
\simeq
C_u(\beta)A r^{\beta-3}\hat{\bf e}_\varphi ,
\label{app:eq:far_field}
\end{equation}
where
\begin{equation}
C_\theta(\beta)
=
\frac{2^{2-\beta}}{2\pi}
\frac{\Gamma\left(\frac{3-\beta}{2}\right)}
{\Gamma\left(\frac{\beta-1}{2}\right)},
\qquad
C_u(\beta)
=
-
\frac{2^{2-\beta}}{2\pi}
\frac{\Gamma\left(\frac{4-\beta}{2}\right)}
{\Gamma\left(\frac{\beta}{2}\right)} .
\label{app:eq:coefficients}
\end{equation}
The stream-function perturbation has the far-field behavior
\begin{equation}
\delta\psi_0(r)
\simeq
C_\psi(\beta)A r^{\beta-2},
\end{equation}
with
\begin{equation}
C_\psi(\beta)
=
\frac{2^{1-\beta}}{2\pi}
\frac{\Gamma\left(1-\frac{\beta}{2}\right)}
{\Gamma\left(\frac{\beta}{2}\right)} .
\end{equation}

The exponent $\beta$ controls the relative amplitude of the algebraic far-field tails of the buoyancy and velocity perturbations, which does not depend on the regularization width $\sigma$. 
The value $\beta=0.4$ was chosen so that these two far-field amplitudes are approximately matched. 
Indeed, the condition
\begin{equation}
C_\theta(\beta^*)=C_u(\beta^*)
\end{equation}
gives
\begin{equation}
\beta^*\simeq0.4064 \ .
\end{equation}
 
Therefore, $\beta$ does not introduce a new length scale; it only fixes the detailed geometry of the localized perturbation. 
We expect the results found in the main document to be robust with respect to the shape of the perturbation, up to geometry-dependent constants, provided that the buoyancy and velocity perturbations have comparable far-field decay.

\begin{figure}[h!]
\centering
\includegraphics[width=0.65\linewidth]{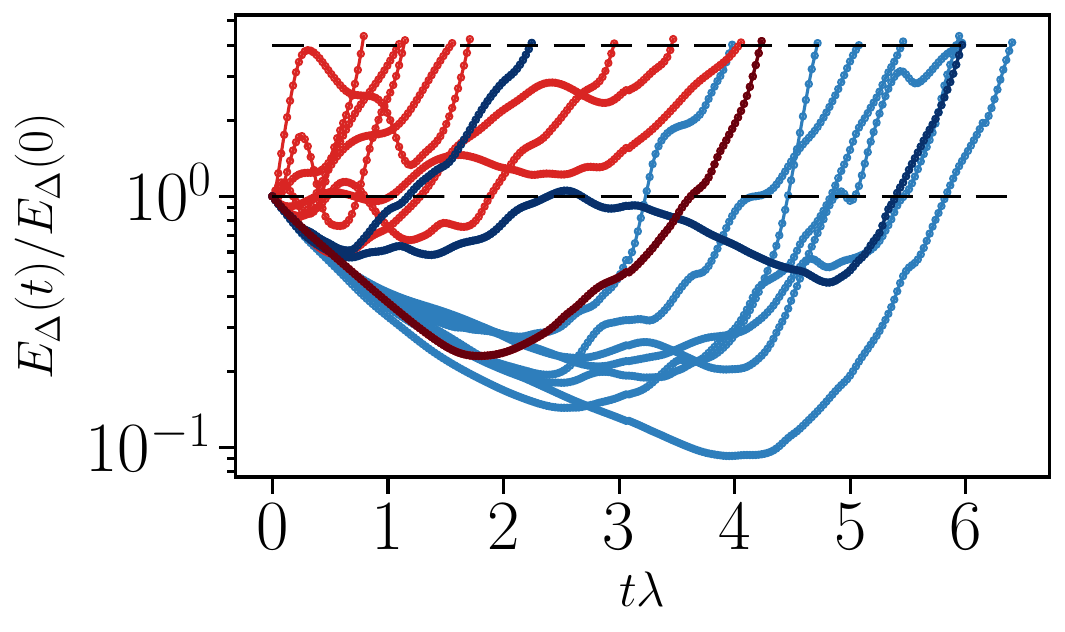}
\caption{
Time evolution of the normalized error energy $E_\Delta(t)/E_\Delta(0)$
for the localized perturbations considered in Fig.~3(a).
The trajectories approximately cluster into two families, corresponding to
perturbations initialized inside (blue) and outside (red) coherent vortical structures.
The darker curves highlight realizations that depart from the typical behavior
of their respective group.
}
\label{fig:energytrajectories}
\end{figure}

\section{Dependence of error growth on the local flow structure}
\label{appendix2}

\ \ \ The return times reported in Fig.~3(a) are associated with markedly different
histories of the error energy. Figure~\ref{fig:energytrajectories} shows the
corresponding trajectories $E_\Delta(t)$ for the same set of localized
perturbations. The trajectories approximately cluster into two families,
consistently with the separation observed in Fig.~3(a). Perturbations
initialized in regions of small $|\theta|$ typically recover their initial energy
rapidly, whereas perturbations initialized inside coherent vortices undergo a
substantially longer transient decay. The variability of the return time
therefore reflects distinct dynamical histories of the perturbation.

A few trajectories depart from this simple classification, as highlighted by
the darker curves in Fig.~\ref{fig:energytrajectories}. Inspection of the
corresponding flow evolution shows that these cases are associated with changes
of the local structure during the perturbation history. In some realizations,
a vortex initially surrounding the perturbation disappears on a relatively
short time scale. These cases also
account for the outliers observed in Fig.~3(a), emphasizing that the return
time cannot, in general, be inferred from the initial value
$\theta(\mathbf{x}_0)$ alone. Rather, it is sensitive to the local structures
experienced by the perturbation during its evolution.

This observation motivates us to isolate the role of local flow geometry.
To this end, we performed short inviscid and unforced simulations, with
$\nu=0$, $\mu=0$, and $f=0$, using two idealized initial conditions
representative of the local configurations encountered in the turbulent flow.
In the first case, the base field $\theta^v(\mathbf{x})$ contains an
approximately uniform circular coherent structure centered near the
perturbation. In the second case, the base field $\theta^s(\mathbf{x})$
contains a straining configuration at the perturbation location. In both
cases, the base field was perturbed using the same localized perturbation
adopted in the main simulations.

These idealized configurations are not intended to reproduce quantitatively
the turbulent dynamics, but rather to isolate the geometric mechanism
suggested by the trajectories in Fig.~\ref{fig:energytrajectories}.

\begin{figure}[h!]
\centering
\includegraphics[width=0.75\linewidth]{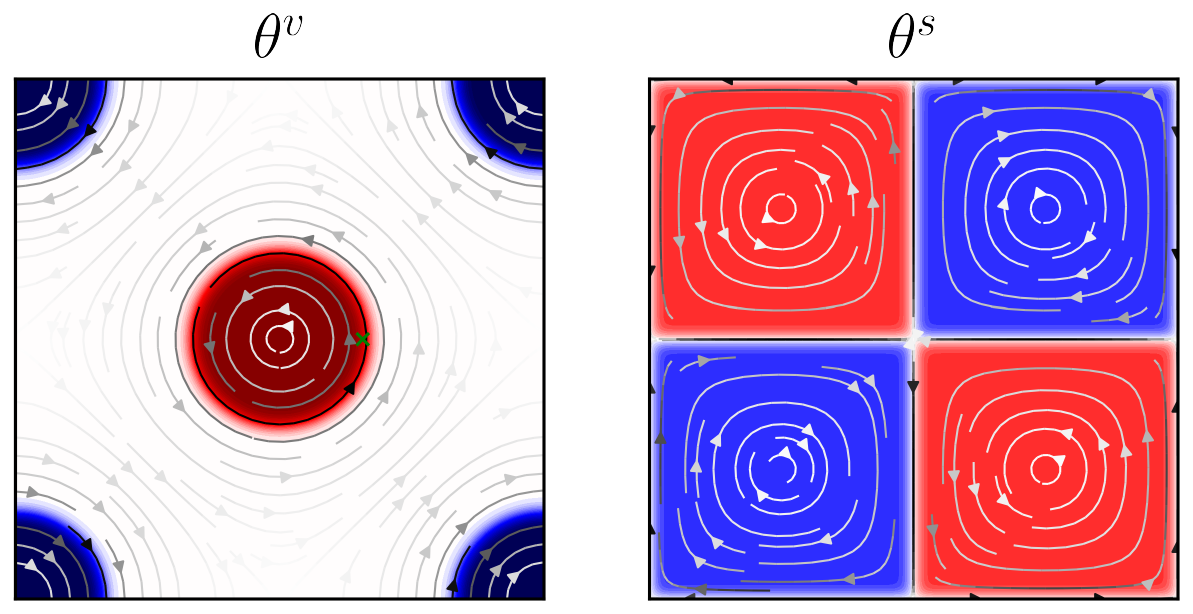}
\caption{
Idealized buoyancy-field configurations used to investigate the geometric mechanism of error growth.
Left: vortical configuration $\theta^v$. The green cross indicates the
perturbation center in the shifted case.
Right: straining configuration $\theta^s$.
Streamlines are superimposed to illustrate the local velocity geometry.
}
\label{fig:supp1}
\end{figure}

We considered three perturbations. In the first, denoted by $\delta\theta^v$,
the vortex field $\theta^v(\mathbf{x})$ is perturbed at a position
$\mathbf{x}_0$ that coincides with the vortex center. In the second, denoted
by $\delta\theta^v_{\mathrm{shift}}$, the perturbation is shifted toward the
edge of the coherent vortex, as indicated by the green cross in
Fig.~\ref{fig:supp1}. Finally, in the third configuration, denoted by
$\delta\theta^s$, the perturbation is centered in the region of strongest
strain in $\theta^s(\mathbf{x})$. In all cases, we use $\sigma=4$ grid points.
The corresponding fields are shown in Fig.~\ref{fig:supp1}.

\begin{figure}[h!]
\centering
\includegraphics[width=0.9\linewidth]{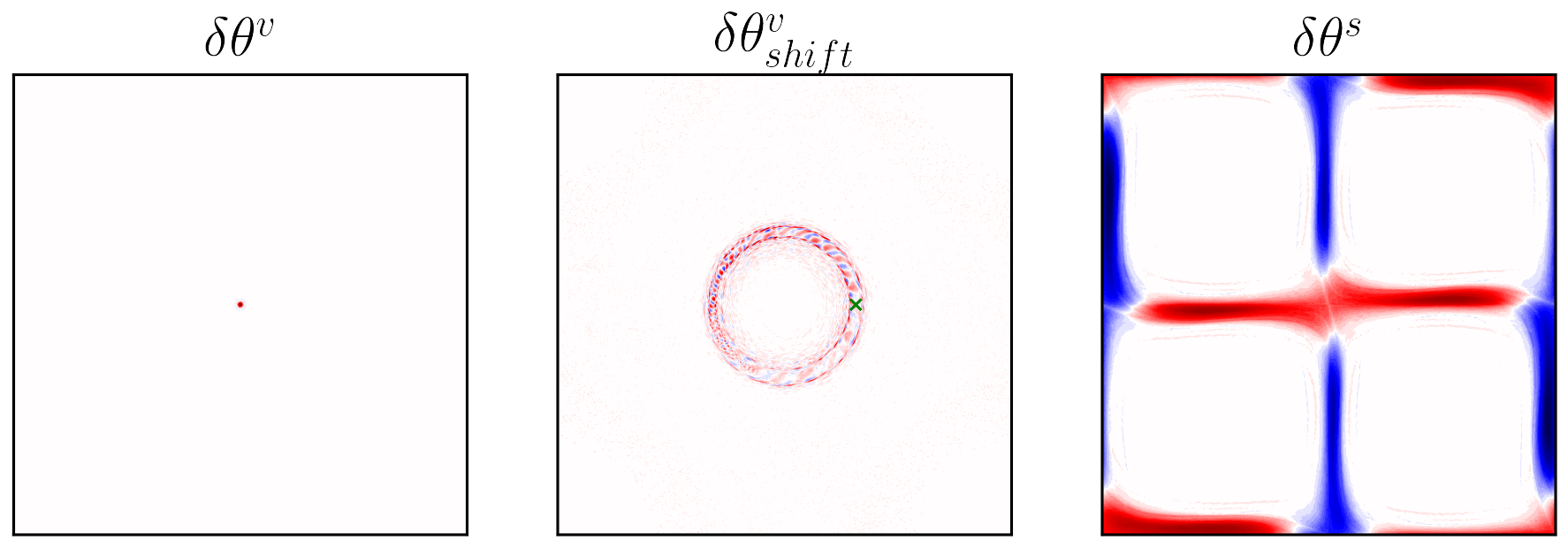}
\caption{
Evolution of the perturbation field in the vortical and straining
configurations. The color scale is normalized independently in each panel.
In the vortical cases, the perturbation remains localized, whereas in the
straining case it is rapidly stretched.
}
\label{fig:supp2}
\end{figure}

In both configurations, the energy $E$ of the base field is normalized to
unity, while the initial perturbation energy is set to
$E_\Delta=5\times 10^{-9}$. Figure~\ref{fig:supp2} shows the perturbation
fields at time $t=20$ (simulation units). In the straining configuration, the
perturbation is rapidly stretched and the error energy increases. By contrast,
in the vortical configuration, the perturbation remains trapped within the
coherent structure. The error energy grows only slowly in the
$\delta\theta^v_{\mathrm{shift}}$ case and does not grow appreciably in the
centered $\delta\theta^v$ case.

\begin{figure}[h!]
\centering
\includegraphics[width=0.9\linewidth]{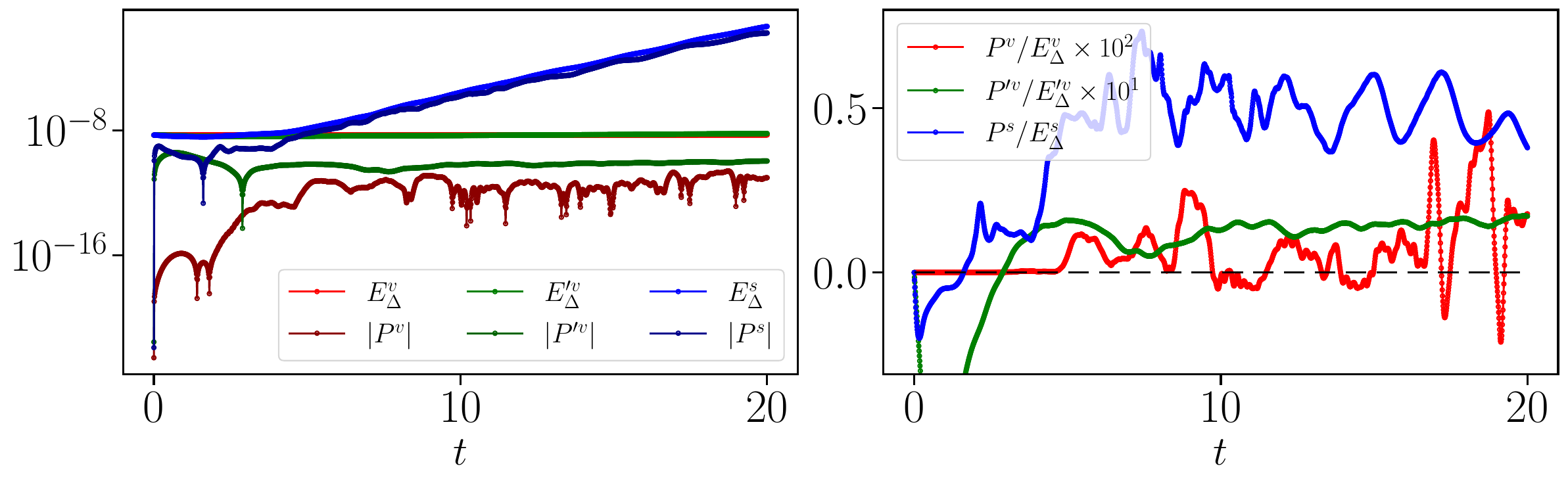}
\caption{
Time evolution of the error energy $E_\Delta(t)$ and the production term
$P(t)$ in the idealized vortical and straining configurations.
The shifted vortex configuration is denoted by a prime.}
\label{fig:supp3}
\end{figure}

In Fig.~\ref{fig:supp3}, the left panel shows the time evolution of the error
energy, $E_\Delta=\langle \delta\theta^2\rangle/2$, together with the
production term,
$P(t)=-\langle \delta\theta\,\delta\mathbf{u}\cdot\nabla\theta\rangle$,
for the three perturbations. In the centered vortex configuration, $P$
remains very small and oscillates around zero. In the straining configuration,
by contrast, the production term becomes positive and reaches substantially
larger values after an initial transient.

Since these idealized simulations are inviscid and unforced, the error-energy
balance reduces to
\begin{equation}
\frac{dE_\Delta}{dt}=P(t).
\end{equation}
Therefore, the ratio $P(t)/E_\Delta(t)$ shown in the right panel of
Fig.~\ref{fig:supp3} is the instantaneous logarithmic growth rate of the
error energy,
\begin{equation}
\frac{P(t)}{E_\Delta(t)}
=
\frac{d}{dt}\log E_\Delta(t).
\end{equation}
In the linear exponential-growth regime, this quantity is expected to be a constant. Apart from fluctuations, an approximately constant logarithmic growth rate is indeed observed after an initial transient. Such a growth rate oscillates around a large value in the straining configuration, being substantially smaller in the $\delta\theta^v_{\mathrm{shift}}$ case, and remains close to zero for the centered $\delta\theta^v$ case (see the right panel of Fig.~\ref{fig:supp3}).

These results support the interpretation that the delayed error growth observed in the main text is mostly related to the local geometry of the flow surrounding the localized perturbation. Perturbations initialized inside coherent vortices experience weak production for short times, because $\delta\mathbf{u}$ is nearly tangent to the local level sets of $\theta$. As long as this geometry persists, the logarithmic growth rate remains small. By contrast, perturbations in straining regions rapidly develop a sustained positive production term and are efficiently amplified.

In the forced-dissipative turbulent system of the main text, diffusion may provide an additional mechanism, particularly for perturbations initialized near the edge of a vortex. By progressively weakening their confinement, diffusion may allow the perturbation to sample nearby straining regions and thereby enhance the production term.

\end{document}